\documentclass{article}

\usepackage{arxiv}

\usepackage[utf8]{inputenc} % allow utf-8 input
\usepackage[T1]{fontenc}    % use 8-bit T1 fonts
\usepackage{booktabs}       % professional-quality tables

\usepackage{nicefrac}       % compact symbols for 1/2, etc.
\usepackage{microtype}      % microtypography
\usepackage{lipsum}
\usepackage{multirow}
\usepackage{setspace}
\usepackage{kantlipsum}
\usepackage{lscape}
\usepackage{amssymb}
\usepackage{graphicx}
\usepackage{upgreek}
\usepackage{amsmath}
\usepackage{amsfonts}
\usepackage{amssymb}
\usepackage{amsthm}
\usepackage{cleveref}

\title{Graph Convolutional Neural Networks for analysis of EEG signals, BCI application}

\author{
  Mirfarid Musavian Ghazani \\
  Skolkovo Institute of Science and Technology \\
  (SKOLTECH), CDISE, Moscow, Russia\\
  \texttt{mirfarid.musavian@skoltech.ru} \\
   \And
 Anh Huy Phan \\
  Skolkovo Institute of Science and Technology \\
  (SKOLTECH), CDISE, Moscow, Russia, \\
}

\begin{document}
\maketitle

\begin{abstract}
Decoding brain signals has gained many attention and has found much applications in recent years such as Brain Computer Interfaces, communicating with controlling external devices using the user's intentions, occupies an emerging field with the potential of changing the world, with diverse applications from rehabilitation to human augmentation. This being said brain signal analysis, EEG brain signal analysis in particular, is a challenging task. With the advances and achievements in the field of deep learning in problem solving with using only raw data, few attempts has been carried in recent years, to apply deep learning to tackle EEG among other types of brain signals. In this study, we propose a novel loss function, called DeepCSP to extend the classical Common Spatial Patterns to a non linear, differentiable module to serve as the loss function to enforce linearly separable latent representations of EEG signals belonging to different classes in an end to end manner on raw signals without the need to perform extensive feature engineering. With recent generalizations of deep learning methods to work on arbitrarily structured graphs and the introduced loss we have proposed two light weight models to decode EEG signals and carried experiments to show their performance.
\end{abstract}

% keywords can be removed

\section{Introduction}
A Brain Computer Interfaces (BCI) is "a combination of hardware and software that allows brain activities to control external devices or even computers". Decoding and classification of mental intents using the data gathered from the brain activities recorded from subjects in a very carefully designed experiment setting is the main goal in these applications.

Among the various data devices and technologies used in recording brain activities, EEG recording modalities are of special interest. In the case of this kind of data, two main parameters are spatial resolution, referring to the spatial proximity of the recorded data point and temporal resolution, referring to the sampling frequency. The high temporal resolution of EEG recordings combined with relatively much lower costs compared to other methods and the fact that it is non invasive, has been the most popular tool in BCI applications and therefore is chosen as the focus of this study. There are variety of applications for BCIs, from control of robotic arms and various other prosthetic devices \cite{mcfarland2008brain} to speech synthesizers \cite{lotte2007review} and other communication devices \cite{guan2004high}. In general.

It can be argued that however attempting to decode the brain signals obtained among the various experiment settings faces lots of challenges, decoding Motor Imagery (MI) signals is relatively more troublesome. 

BCI's can be defined as "a combination of hardware and software that allows brain activities to control external devices or even computers"\cite{ramadan2017brain}. It can be argued that one of the most challenging and crucial parts of these interfaces, is designing the experiments to record data from brain. Motor Imagery (MI) is defined as "mental rehearsal of movements without the movements being executed" \cite{spence2013crossmodal}. An example of MI can be the intent of hand movement versus the physical act of hand movement. When movement is imagined, the activity in the brain is in the motor cortex, like the real movement.

Despite vast and successful applications and developments in deep learning in recent years and their generalizations to Graph Convolutional Neural Networks, the amount of studies that have focused on the BCI Motor Imagery applications are limited. 

In this study, we develop a new loss function, DeepCSP, inspired by the ubiquitous Common Spatial Patterns method used in BCI application. The applicability of non linear extension of this method and the effect of building a differentiable module of one of the most commonly used signal processing, on the adjustment and optimizing the parameters of the deep convolutional architectures to get separable latent features of EEG signals can be power full addition to end to end BCI applications.

By representing EEG signals as graphs we investigate the positive effect of the features learned from GCNs, powerful paradigm linking deep learning to analysis and classification of various types of graphs, on the performance of the Motor Imagery EEG signals classification. 

\section{Literature Review}

In BCI systems, machine learning has a major role in the classification of EEG signals. Supervised algorithms such as SVM (with different kernels) \cite{kang2009composite, arvaneh2011optimizing}, LDA \cite{samek2012stationary, arvaneh2013eeg} and Bayesian methods \cite{arvaneh2011optimizing} have been among the most popular algorithms, mainly due to their simplicity and fast training \cite{lotte2007review}. One main drawback of these methods is that they require data preprocessing and manual feature extraction in order to be able to learn the data.

Most of the published techniques using the mentioned class of machine learning algorithms use some form of feature extraction followed by classification with a relatively simple model. This fact causes these techniques to be fast and relatively less likely to suffer from overfitting, in the meantime their accuracy and efficiency heavily depend on the quality of the feature extraction methods used. They function by transforming the training data to a space, mainly such that the different classes are more easily distinguishable and then find a hyperplane in that space that tries to separate the classes of data. The Linear Discriminant Analysis (LDA) is one of the commonly used classifiers of this group of approaches which assumes the normality and homoscedasticity (meaning that the classes have the same variance) to create a linear decision surface \cite{xu2011enhanced}. One of the other commonly used classifiers is Naive Bayes, a simple probabilistic classifier which uses the concept of maximum likelihood to make decisions \cite{machado2013study}, and K Nearest Neighbors, which queries the training data using a distance metric to find the K samples closest to the one being classified to determine the label \cite{lotte2007review}.
The most common feature extraction methods include: Fourier transforms and its variants, wavelet transforms, principal component analysis (PCA), independent component analysis (ICA), autoregressive methods, or combinations of those techniques. The Fourier transform is likely the most common feature extraction method. It provides information on the frequency spectrum of the signals, extracting the spectral information but failing to capture the temporal information of the signals. One of the methods used to overcome this shortcoming heavily deployed in the literature is to use the short time Fourier transforms. The Fourier transform is used to calculate features, such as power spectra, which match nicely with EEG literature on brain activities that occur within different frequency bands \cite{li2009emotion, lotte2007review}.

As mentioned before new approaches to tackle the challenge decoding EEG signals, and minimizing the need to perform feature engineering steps on the signals, some of the recently published studies have attempted to use deep learning architectures, while maintaining the competitive performance levels.  

One of the early introduced deep learning architectures presenting successful results at classifying Steady state visually evoked potential, SSVEP signals, used series of Spatial convolutional layers to mix the input EEG signals, followed by 1D temporal convolutional layers  \cite{cecotti2011time}. before feeding the extracted features from these two layers, the features have been switched into the frequency domain using fast Fourier transform. 

In another study both the raw form of the EEG signals and extracted spectral features have been used as input for doing classification on the dataset of recorded EEG signals of subjects, while imagining fragments from a span of music, selected from different \cite{stober2014using}. The results showed the capacity of CNNs to classify imagery based EEG. In a yet another study, the same authors employed a convolutional autoencoder (CAE) to train a CNN on the same dataset, using cross trial similarity constraint encoding in order to learn individually adapted components that are shared in between the subjects, in order to capture invariance between trials within and across subjects \cite{stober2015deep}. The use of autoencoders in this study is one of the earliest successful attempts to use deep learning architectures to do data augmentation to tackle the issue of the low number of samples of EEG data.

A novel representation of EEG signals, where the channels of the EEG signals were transformed into two dimensional space and then by using Fourier transform across several time bins, the data was reduced to fewer frames \cite{bashivan2015learning}. By doing so, a sequence of images is generated for the whole EEG trial and then fed into a combined CNN and Long Short Term Memory (LSTM), and then a fully connected layer was used for classification. This method was able to outperform the classical methods by considerable margins.

In \cite{lawhern2018eegnet} use of fully convolutional neural networks was explored by Lawhern et al, on a variety of BCI paradigms, including P300 visual evoked potentials, error related negativity responses (ERN),
movement related cortical potentials (MRCP), and sensory motor rhythms (SMR). The network proposed in this study includes temporal convolution layer to learn frequency filters, then a depth wise convolution layer, connected to each feature map individually, enabling the model to learn frequency specific spatial filters, followed by a separable convolution layer, which attempts to learns a temporal summary for each feature map individually, followed by a point wise convolution, which learns how to optimally mix the feature maps together. The achieved state of the art results of this study in many BCI paradigms can be linked to the low number of parameters of the model and well designed architecture employed.

In \cite{santana2014joint} proposed a neural network based on the CSP algorithm, consisting of three steps, first a temporal convolutional for each channel, then a spatio temporal projection of the filtered signal, which is implemented as a single fully connected feedforward layer or two divided layers, one for spatial projection and another for temporal projection. Finally, the extracted features are fed into the classification layer. The weights of the spatial layer are initialized by CSP. In a similar paper, \cite{yuksel2015neural} also focused on optimizing a neural network for spatial filter optimization by using only two fully connected layers.

In the paper \cite{yang2015use}, an augmented CSP (ACSP) algorithm is suggested by using overlapping frequency bands. to extract features and use them as an input for the CNN. This study suffers from the need to do a fair amount of manual feature engineering and even does not take into account the temporal characteristics of the EEG signals which can be accounted for relatively poor results achieved in the motor imagery paradigm

\cite{tabar2016novel} used a stacked autoencoder (SAE) and a CNN to classify the two class, left/right hand
MI task, three electrodes C3, Cz and C4, which are located on the motor area of the brain. The signals from these selected electrodes are transformed using a short time Fourier transform (STFT). The output of the convolutional layer is passed to SAE and then to a classifier. This study has outperformed SVM. 

\cite{schirrmeister2017deep} proposes and experiment with three deep learning architectures, Deep ConvNet, Shallow ConvNet, and Hybrid Convnet for classifying motor imagery signals and outperforms the accuracy of the widely used filter bank common spatial patterns (FBCSP) with minimal preprocessing applied to EEG signals. Deep ConvNet consists of four convolution max pooling blocks, with a special first block designed to handle EEG input, followed by three standard convolution max pooling blocks and softmax classification layer. Shallow ConvNet architecture consists only of two layers, one temporal convolutional layer with bigger kernel size compared to Depp ConvNet and then a spatial convolutional layer, followed by standard non linearity and mean pooling layer. Hybrid ConvNet just stacks both networks and fuses the features learned from them after the final layer. The results of this study show that the shallow networks if designed in an efficient manner, can achieve better results than more deep models and even standard classifiers used in the BCI applications.

Few attempts have been carried in the literature to apply Graph Convolutional architectures to decode brain signals, even fewer attempts have been aimed at BCI motor imagery applications. 
In \cite{8758787}, authors used phase locking value (PLV) connectivity estimation to construct the graph and used manually extracted time frequency domain feature univariate features, including the power spectral density (PSD) feature, the differential asymmetry (DASM) feature, the rational asymmetry (RASM) feature and the differential caudality (DCAU) features in order to decode the various emotional states of the subjects. The architecture employed consists of a graph convolutional layer, two ReLu layers, and a fully connected layer followed by the softmax layer to perform the classification. The published results in this study, outperforms SVM, Deep Belief Networks and GCNN.

In another study \cite{zhang2020motor}, the EEG signals have been represented as graph based on the euclidean distance of the electrodes in sensor space. This matrix has been used to embed the signals in a CSP fashion and then the embedded graph has been cut into several temporal slices. Two models has been introduced to extract features from these slices, one CNN model including a convolutional layer followed by a MaxPooling layer and the other architecture using Long Short Term Memory (LSTM) units to explore the temporal attentive temporal dynamics of the slices. These introduced models achieved superior results in decoding Motor Imagery EEG signals.

In s very recent study \cite{hou2020deep}, authors have proposed a novel architecture combining the BiLSTM with the GCN to decode EEG tasks which used Pearson Correlation as an estimate for the graph adjacency matrix, reporting to have achieved the state of the art performance on Motor Imagery EEG classification on Physionet dataset.

\section{Proposed Methods}
\subsection{DeepCSP, novel EEG tailored loss} \label{DeepCSP}

Inspired by \cite{dorfer2015deep}, where the authors introduced the formulation of DeepLDA version of Linear Discriminant Analysis (LDA) enabling the deep learning networks to impose a separation of classes in the latent space produced by the network by maximizing the ratio of the between and within class scatter by means of end to end backpropagation, we introduce the same concept to Common Spatial patterns (CSP).

CSP is the most common method used in motor imagery signals to achieve accurate spatial features and easily distinguishable classes. Motor Imagery experiment deals with predicting the intent of the subjects, commanded to them by a cue, such as on the screen. In this study, we only investigate classifying two class cases. 

Let $C_1$ and $C_2$ be different intents, imagined by the subjects, for example imagining left hand vs right hand movement. Each trial in which subjects are requested to perform a task, is called an epoch. The EEG signal recorded for each epoch can be represented as $\mathbf{X} \in \mathbb{R}^{D \times T}$, $D$ denoting electrodes (channels), and $T$ the number of time points sampled in each trial. $y \in \{0, 1\}$ being the class labels, the objective of the classification is to learn the relations in $ \{ \mathbf{X} _{i},y_{i} \}_{i=1}^{N}$, where $N$ is the total number of training trials, such that it can perform well predicting the unseen trials in test sets.

CSP method reduces the dimensionality of each trial by extracting spatial components that are common to both classes, in a way that the signals after being projected onto these components have high variance in between the classes. This method uses simultaneous diagonalization of the covariance matrices of the classes.
\noindent
Covariance matrices for trials belonging to each class. $C_1$ and $C_2$ are given by:
\begin{equation} \label{eq:csp1}
    C_{1}={{(X_{1}X_{1}^{T})}\over{trace(X_{1}X_{1}^{T})}}
\end{equation}

\begin{equation} \label{eq:csp2}
    C_{2}={{(X_{2}X_{2}^{T})}\over{trace(X_{2}X_{2}^{T})}}
\end{equation}
\noindent
Then these covariances are averaged to get averaged normalized covariance matrices $\overline{C_{1}}$ and $\overline{C_{2}}$, and add to obtain the composite covariance matrix C.

\begin{equation} \label{eq:csp3}
    `C = \overline{C_{1}} + \overline{C_{2}}
\end{equation}
\noindent
Since $C$ is symmetric, its eigen decomposition has the form

\begin{equation} \label{eq:csp4}
    \mathbf{C} = \mathbf{U} {\mathbf{\Lambda}} \mathbf{U} ^{\top}
\end{equation}
\noindent
from which we can obtain the whitening transformation matrix $P$ as follows:

\begin{equation} \label{eq:csp5}
    \mathbf{P} = {\mathbf{\Lambda}} ^{- {{1}/ {2}}} \mathbf{U} ^{\top}
\end{equation}
\noindent
The transformed covariance matrices ${\mathbf{\Sigma}} _{1}= \mathbf{P} \mathbf{C} _{1} \mathbf{P} ^{\top}$ and ${\mathbf{\Sigma}} _{2}= \mathbf{P} \mathbf{C} _{2} \mathbf{P} ^{\top}$ share common eigen vectors and the sum of them has the following property:

\begin{equation} \label{eq:csp6}
    {\mathbf{\Sigma}} _{1} + {\mathbf{\Sigma}} _{2} = \mathbf{I} _{D \times D}.
\end{equation}
\noindent
The spatial projection matrix $W$ is
\begin{equation} \label{eq:csp7}
    \mathbf{W} = \mathbf{P}  \mathbf{V}^{\top }
\end{equation}

Due to the equation \cref{eq:csp6} top $n$ eigenvectors of ${\mathbf{\Sigma}} _{1}$, when sorted in decreasing order, both account for the maximum variance in the first class and also the minimum variance in the second class and vice versa. Studies shows that $n = 2$ or $n = 3$ helps to achieve good empirical performance \cite{muller1999designing}. The CSP features of the EEG data are computed as
\begin{equation} \label{eq:csp8}
    \mathbf{Z} = \log \left (\mathbf{W}_{2n}^{\top} \mathbf{X} \mathbf{X}^{\top} \mathbf{W}_{2n} \right)
\end{equation}
\noindent
We define the new loss objective as 

\begin{equation} \label{eq:csp9}
    J(W) = {{trace(\mathbf{W}_{2n}^{\top} C_{1} \mathbf{W}_{2n})\over{trace(\mathbf{W}_{2n}^{\top} \left (C_{1} + C_{2} \right) \mathbf{W}_{2n})}}}
\end{equation}

\noindent
imposing separation onto the latent representation over the output of the hidden layers of the deep learning architecture of choice, by not only learning the spatial filters but also optimizing the parameters of the network to ensure more separation using backpropagation.

\subsection{Proposed shallow networks using DeepCSP loss}

As discussed, the results of \cite{schirrmeister2017deep} shows that shallow networks can outperform classic methods like FBCSP and deep network architectures in BCI motor imagery applications.

\begin{figure}
  \includegraphics[width=\linewidth]{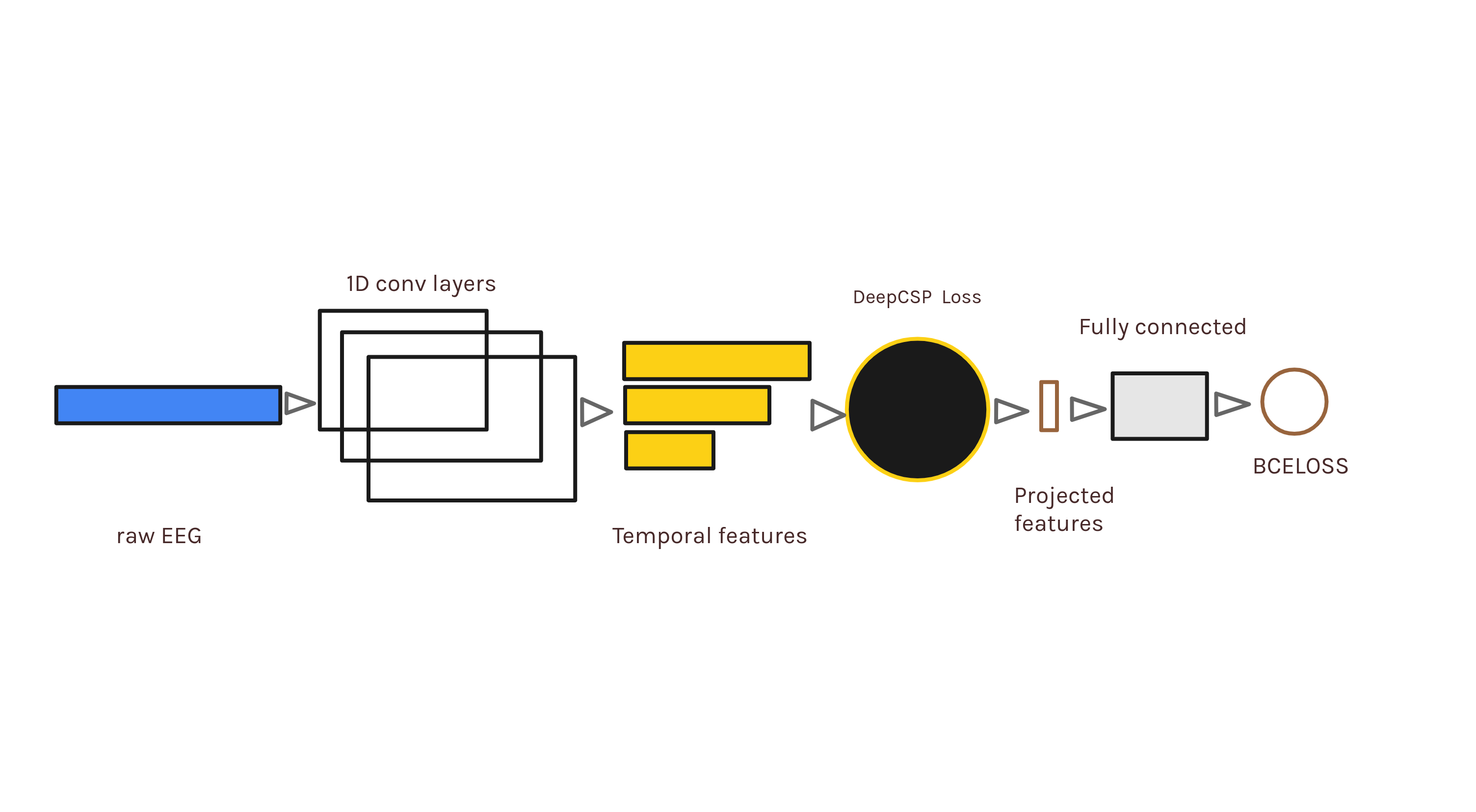}
  \caption{Shallow CNN with DeepCSP loss}
  \label{fig:sdcsp}
\end{figure}

We propose a light weight model composed of two blocks. First block is composed of 3 layers of 1D temporal convolutional filters with multi scale temporal filters, with kernel sizes of respectively half, one third and one forth of the sampling frequency of the devices used in the recording of EEG signals, in an attempt to extract temporal patterns of the input data. The output of these layers are concatenated and fed into the DeepCSP to obtain the spatial filters and reduce the dimension of the output of temporal block further as discussed in \ref{DeepCSP}. Second block consists of two layers of fully connected layers to do classification. A sketch of this model can be found in \cref{fig:sdcsp}

\subsection{Proposed shallow GCN networks using DeepCSP loss}

As mentioned, although combining CNN with graph theory, the Graph Convolutional Neural Network (GCN) has shown competitive results in classifying EEG signals, but very few research has been done to utilize these architectures in MI BCI applications. One main issue in using these class of models is not having a clear cut consensus on how to represents the EEG signals as graphs, in other words the connections and relationship in between the channels. 

In an attempt to address this issue, we investigate two approaches. First we explore the effect of constructing the graphs from EEG signals based on statistical methods developed in graph theoretical analysis of EEG signals. Then we explore constructing the graph dynamically.

\subsection{Statistical methods to construct the graph from EEG}

Here we list the methods widely used in the literature in estimation of various connectivity in between the channels, that we have utilized to construct the graph from EEG signals.

\begin{itemize}
    \item \textbf{Coherence} (Coh) is one of the most commonly utilized connectivity estimators; it is a measurement of the linear relationship of two signals at a specific frequency \cite{nolte2004identifying}.
    \noindent
    Given two time series $x$ and $y$, coherence is given by:
    \begin{equation} \label{eq:graph1}
        coh^2_{xy}(f) = \frac{ |G_{xy}(f)^2| }{ G_{xx}(f) G_{yy}(f) }
    \end{equation}
    \noindent
    where the $G_{xy}(f)$ is the estimated cross spectral density between $x$ and $y$, while $G_{xx}(f)$ and $G_{yy}(f)$ are the autospectrum of $x$ and $y$ respectively.
    \noindent
    The result is a symmetric matrix of size $[n\_channels \times n\_channels]$ bearing no information about the directionality of the interaction, with values within the range $[0,1]$. 
    \item \textbf{Directed Phase Lag Index} (dPLI) was introduced in \cite{stam2012go} to capture the phase and lag relationship as a measure of directed functional connectivity.
    \item \textbf{Phase Locking Value} (PLV) is one of the pioneer methods introduced in \cite{lachaux1999measuring}; it utilizes the Hilbert representation of an EEG time series and quantifies their interaction based on their instantaneous phase in a specific band frequency. 
    \item \textbf{Imaginary part of Phase Locking Value} (IPLV) was proposed to resolve PLV’s sensitivity to volume conduction and common reference effects.
    \noindent
    IPLV is computed similarly as PLV, but taking the imaginary part of the summation \cite{sadaghiani2012alpha}.
    \item \textbf{Phase Lag Index} (PLI) \cite{stam2007phase} proposed as an alternative (to PLV) phase synchronization estimator that is less prone to the effects of common sources (namely, volume conduction and active reference electrodes). These effects can artificially generate functional connectivity as the same signal signal is measured at different electrodes \cite{hardmeier2014reproducibility}. PLI estimates the asymmetry in the distribution of two time series’ instantaneous phase differences.
    \item \textbf{Weighted Phase Lag Index and Debiased Weighted Phase Lag Index} (wPLI) and (dwPLI) were introduced in \cite{vinck2011improved} to achieve better estimations 
    
\end{itemize}

After defining the connectivity between channels, we use it as prior in form of the weighted adjacency matrix and use a layer of GraphSage convolution \cite{hamilton2017inductive} to extract graph features and concatenate them with the temporal features as in \cref{fig:sdcsp}. Our experiments show that defining the temporal features as node feature inputs to GraphSage yields better results.

\section{Results and Discussion}
In order to test the validity of proposed methods, we have used two Motor Imagery datasets and compared the results with some of the state of the art deep learning architectures. First, we give a brief overview of the datasets that we have run the models, then compare the results subject dependent classification results then analyze the proposed methods.

\subsection{Datasets}

The first Motor Imagery dataset is the lla dataset from BCI competition 4 \cite{tangermann2012review}. This dataset consists of EEG recordings of 9 subjects. Four cue based BCI paradigms in these experiments is the imagination of left hand, right hand, both feet and tongue. 12 trials were conducted for each class, in 6 runs separated by short breaks on two different days, a total of 576 trails. We have selected only the trials belonging to the intent of the left hand vs right hand movement of this dataset, since at this point the DeepSCP only supports two classes. 

Each trial was started by a fixation mark on a computer screen for the subjects along with a short warning sound. after 2 seconds subjects were cued by an arrow to perform the associated intent depending on the direction of the arrow, which stayed on the screen for 1.25 seconds, without any feedback. The subjects were asked to repeat the intent until the fixation mark on the screen disappeared. The signals were sampled with. $250 Hz$ and bandpass filtered between $0.5 Hz$ and $100 Hz$. An additional $50 Hz$ notch filter was enabled to suppress line noise.

The second Motor Imagery dataset \cite{steyrl2016random}, consisted of eight runs, five of them for training and three with feedback for validation each run having 20 trials, 50 trials per class for training and 30 trials per class for validation. Subjects were given the task of performing a cued motor imagery (MI) of the right hand and of the for 5 seconds. The Experiment has been carried as the one mentioned in the first dataset, with a sampling rate of $512 Hz$. The electrodes placement was designed for obtaining three Laplacian derivations. Center electrodes at positions C3, Cz, and C4 and four additional electrodes around each center electrode with a distance of 2.5 cm, 15 electrodes total. The participants were aged between 20 and 30 years, 8 naive to the task, and had no known medical or neurological diseases.

\section{Experiments and Results} \label{results}

We have used the ShallowConv and DeepConv CNN's proposed in \cite{schirrmeister2017deep} and EEGNet fourth version \cite{lawhern2018eegnet}. We have used the official implementations provided by the authors and used the recommended training settings by the authors for each one. The optimizer being pytorch implementation of AdamW with $0.00625$ learning rate and zero wight decay and for the DeepConv Net the learning rate of $0.01$ and weight decay of $0.0005$ with a batch size of 64 and Adam optimizer with a learning rate of $0.001$. We have used the EEGNet implementation provided in the BrainCode, a deep learning toolbox to decode raw time domain EEG \footnote{https://braindecode.org/}.

We have trained the Shallow CNN DeepCSP and Shallow GCN with DeepCSP with separate SGD optimizers, one with a learning rate of $0.01$ for updating the parameters of the CNN and GCN part with respect to proposed DeepSCP loss, and one with $0.1$ for updating the parameters for the linear layers of the classification block in consecutive order. Our experiments showed that setting the number of components of the spatial filters learned by the loss to 4, yields better results. So here we include only those experiments. One thing worth mentioning here is that in training, we feed the whole training set to get more accurate spatial filters. Small scale model and data set makes this doable even when not training the model on GPU.

\begin{table}
  \caption{Comparison of results on the BCI competition dataset \cite{tangermann2012review}}
  \label{tbl:BCI1}
  \includegraphics[width=\linewidth]{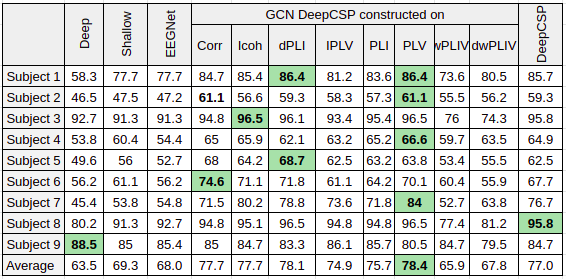}
\end{table}

\begin{table}
  \caption{Comparison of results on the Motor Imagery dataset \cite{steyrl2016random}}
  \label{tbl:BCI2}
  \includegraphics[width=\linewidth]{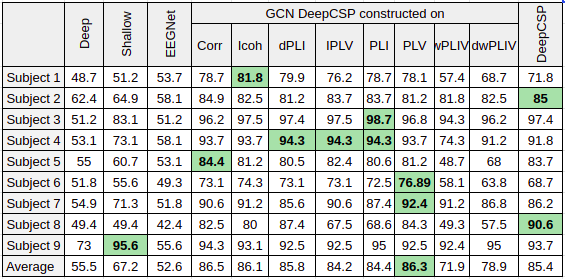}
\end{table}

The results presented in \cref{tbl:BCI1} and \cref{tbl:BCI2}, the proposed DeepCSP adds considerable amount to the accuracy of the classification. Using graph based extracted features, increases the accuracy a small amount too. This comparison of the different approaches to define connectivity matrices of EEG signals and represent them as graphs, effects the model accuracy dramatically. We confirm the findings in \cite{8758787}, that PLV method for connectivity estimation, is a good place to start.

\begin{figure}
  \includegraphics[width=\linewidth]{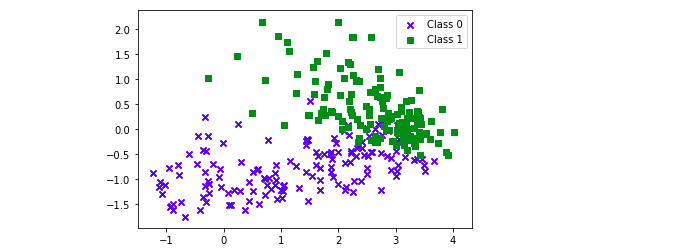}
  \caption{The output of the temporal layer in Shallow DeepCSP after being projected by the spatial filters learned by DeepCSP loss for subject number 3 in BCI competition dataset from \cite{tangermann2012review} . $C_0$ and $C_1$ denotes the EEG signals of intent of right hand and left hand movement, respectively. For visualization reasons we have chosen number of components of spatial filters as two}
  \label{fig:scatter}
\end{figure}

One explanation for these increases in the accuracy can be found if we take a look at the output features of the temporal filters in Shallow DeepCSP. In order to see this, in \cref{fig:scatter}, we plot the transformed outputs of the temporal block, the number of components is set to two. The separation of the classes can be easily noticed, which makes it each for the network to optimize the parameters of the classification block to perform classifying. One has to keep in mind that the spatial filters are learned in the training and then is used to transform the unseen trials.

\begin{figure}
  \includegraphics[width=\linewidth]{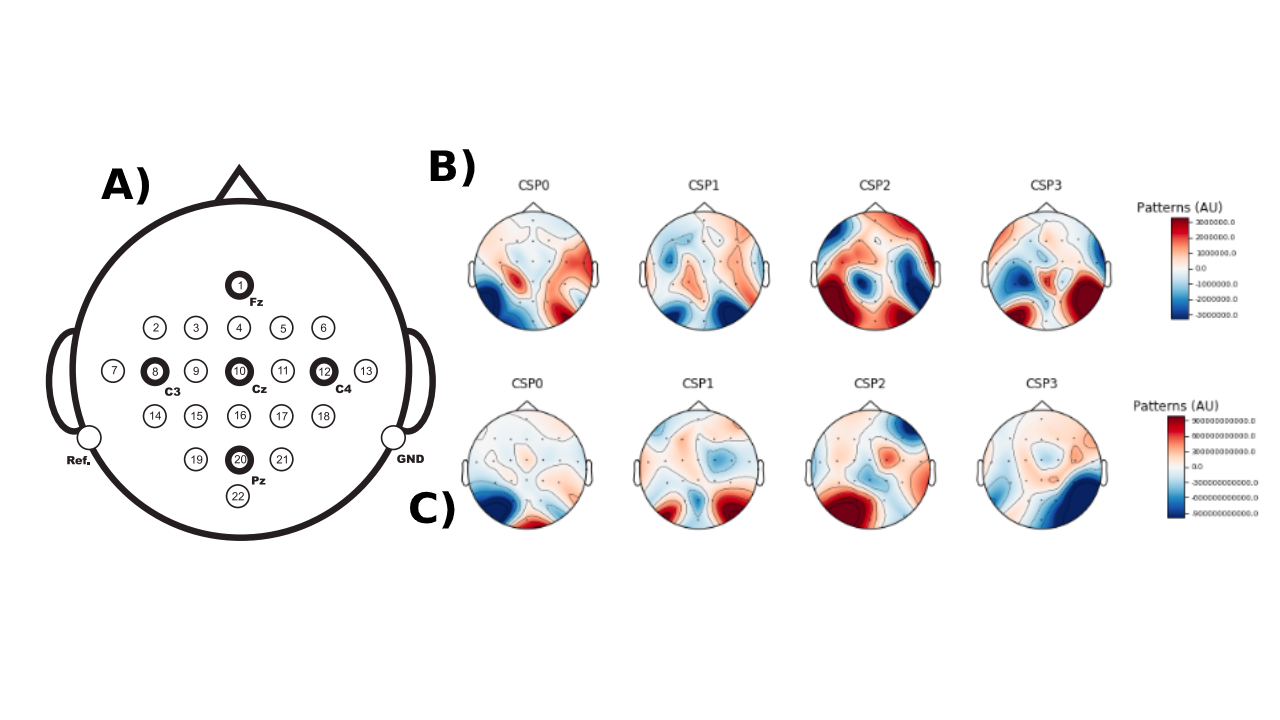}
  \caption{A) Position of electrodes in \cite{tangermann2012review} 
  B) Topomaps generated from spatial filters obtained from DeepCSP
  C) Topomaps generated from spatial filters obtained from CSP }
  \label{fig:filters}
\end{figure}

The spatial filters learned from DeepCSP loss, can be used to plot EEG topographic maps (topomap) to find the brain activation. In the case of performing intent of right or left hand movement, we should see activation in the motor cortex region of the brain which corresponds to electrodes $C3$ and $C4$. "An event with the high feature for the first component (channel C3) and low feature for the second component (channel C4) relates to the right hand imagery movement. On the contrary, this event relates to the left hand imagery movement" \cite{phan2010tensor}.
The plot for the component corresponding to 0 in the far left of section C of \cref{fig:filters} fails to illustrate this. Mainly because one of the shortcomings of CSP method is it requires carefully chosen filter bands. On the other hand, we can clearly see two patterns with different signs over regions of $C3$ and $C4$ in the Topomaps obtained from DeepCSP in an end to end manner.

The results for our limited attempts to employ dynamic graph convolutional networks did not yield good results and have not been presented in here. The reason for the poor results can mainly be contributed to the large increase in the number of parameters in the architecture to update the graph dynamically at each iteration, which causes the model to over fit very quickly in early epochs, which is mainly due to the large the noise to signal ratio in the EEG signals.   

\section{Conclusion and future research plans}
Our preliminary experiments provided in the previous section, the proposed non linear, differentiable generalization of common spatial filters as a new loss, DeepCSP, shows promising results in Motor Imagery BCI applications by forcing separation on the output feature space of the CNN and GCN layers. Despite the good performance of the model proposed, it is designed for two class experiments. This can be solved by extending the DeepCSP for multi class case, benefiting from the existing literature of CSP method for multi class settings.
We have seen that representing EEG signals as graphs and extracting graph features using recent graph convolutions, adds to the performance of the model. However, it is noteworthy to mention that among various kinds of graph convolutions available, we have only experimented with GraphSage, only because of the availability of the source code and ease of use of it.  The main goal of future research is to extend this model to subject independent applications. 

\bibliographystyle{unsrt}

\end{document}